\pdfoutput=1

\documentclass[%
 reprint,
 superscriptaddress,
 amsmath,amssymb,
 aps,
 prl,
floatfix,
]{revtex4-2}

\usepackage{graphicx}
\usepackage{dcolumn}
\usepackage{bm}
\usepackage[euler]{textgreek}%
\usepackage{xcolor}

\begin{document}

\preprint{APS/123-QED}


\title{
Near-Quantum-Noise Axion Dark Matter Search at CAPP around 9.5 $\mu$eV \\
}

\author{Jinsu Kim} 
\affiliation{Department of Physics, KAIST, Daejeon 34141, Republic of Korea}\affiliation{Center for Axion and Precision Physics Research (CAPP), IBS, Daejeon 34051, Republic of Korea}

\author{Ohjoon Kwon} \affiliation{Center for Axion and Precision Physics Research (CAPP), IBS, Daejeon 34051, Republic of Korea}

\author{\c{C}a\u{g}lar Kutlu} \affiliation{Department of Physics, KAIST, Daejeon 34141, Republic of Korea}\affiliation{Center for Axion and Precision Physics Research (CAPP), IBS, Daejeon 34051, Republic of Korea}

\author{Woohyun Chung} 
 \email{Corresponding author. \\ gnuhcw@ibs.re.kr}
 \affiliation{Center for Axion and Precision Physics Research (CAPP), IBS, Daejeon 34051, Republic of Korea}

\author{Andrei Matlashov} \affiliation{Center for Axion and Precision Physics Research (CAPP), IBS, Daejeon 34051, Republic of Korea}

\author{Sergey Uchaikin} \affiliation{Center for Axion and Precision Physics Research (CAPP), IBS, Daejeon 34051, Republic of Korea}

\author{Arjan Ferdinand van Loo} \affiliation{RIKEN Center for Quantum Computing (RQC), Wako, Saitama 351–0198, Japan}\affiliation{Department of Applied Physics, Graduate School of Engineering, The University of Tokyo, Bunkyo-ku, Tokyo 113-8656, Japan}

\author{Yasunobu Nakamura} \affiliation{Department of Applied Physics, Graduate School of Engineering, The University of Tokyo, Bunkyo-ku, Tokyo 113-8656, Japan}

\author{Seonjeong Oh} \affiliation{Center for Axion and Precision Physics Research (CAPP), IBS, Daejeon 34051, Republic of Korea}

\author{HeeSu Byun}\affiliation{Center for Axion and Precision Physics Research (CAPP), IBS, Daejeon 34051, Republic of Korea}

\author{Danho Ahn} 
\affiliation{Department of Physics, KAIST, Daejeon 34141, Republic of Korea}\affiliation{Center for Axion and Precision Physics Research (CAPP), IBS, Daejeon 34051, Republic of Korea}

\author{Yannis K. Semertzidis}
\affiliation{Center for Axion and Precision Physics Research (CAPP), IBS, Daejeon 34051, Republic of Korea}

\affiliation{Department of Physics, KAIST, Daejeon 34141, Republic of Korea}

\date{\today}


\begin{abstract}
We report the results of an axion dark matter search over an axion mass range of 9.39--9.51~\textit{μ}eV. A flux-driven Josephson parametric amplifier (JPA) was added to the cryogenic receiver chain. A system noise temperature of as low as 200~mK was achieved, which is the lowest recorded noise among published axion cavity experiments with phase-insensitive JPA operation. In addition, we developed a two-stage scanning method which boosted the scan speed by 26\%. As a result, a range of two-photon coupling in a plausible model for the QCD axion was excluded with an order of magnitude higher in sensitivity than existing limits.
\end{abstract}

\maketitle



Axions, hypothetical particles associated with the spontaneous breaking of a postulated global U(1) symmetry, offer a dynamic solution to the strong CP problem, an important puzzle in the standard model (SM) \cite{PQ}. Axions in the mass range of 1 $\mu$eV -- 10 meV are considered a favored candidate for dark matter \cite{RevModPhys.75.777,refId0}, but they have extremely weak interactions with the SM fields \cite{KSVZ1979,*KSVZ1980, *KSVZ1987, DFSZ_Zhitnitsky1980, *DFSZ1981} which makes relevant searches exceptionally difficult. To date, Sikivie’s haloscope remains the most sensitive approach in this mass range \cite{Sikivie}. Relying on the two-photon coupling that converts an axion into a single photon, it utilizes a frequency-tunable microwave cavity immersed in a strong magnetic field. 

One of the most effective tools for improving the haloscope's performance is the quantum-noise-limited amplifier, which was successfully developed over the past decade, and has already been used to increase the sensitivity of many axion search experiments \cite{PRL_ADMX2010, brubaker2017first}. The Center for Axion and Precision Physics Research (CAPP) recently implemented a flux-driven Josephson parametric amplifier (JPA) in the receiver chain of the axion haloscope, which reduced the noise of the system to the quantum regime. The flux-driven JPA is a quantum-noise-limited amplifier consisting of a superconducting quantum interference device (SQUID) at the end of a coplanar waveguide $\lambda /4$ resonator \cite{flux_jpa}. A DC magnetic flux through the SQUID loop tunes the resonant frequency of the JPA, while an inductively coupled pump tone is used to operate it in the three-wave mixing mode, which induces a parametric amplification \cite{capp_jpa, uchaikin2019development}. While the JPA's limitations in instantaneous bandwidth makes it difficult to realize its full potential for CAPP's experimental programs, its contribution to improving the overall system noise temperature is immense.

In this letter, we report the recent results of a Pilot Axion-Cavity Experiment (CAPP-PACE), where a JPA was integrated into a conventional axion dark matter experiment to obtain the full benefit of a quantum-noise-limited amplifier. With the upgraded receiver chain the experiment recorded the lowest noise among published axion haloscope searches under phase-insensitive operation, improving frequency scan speed by a factor of 50. We also introduced a newly developed scanning method which statistically optimizes the total scan time by dividing the scanning process into two steps. The new method has greatly improved the net scan speed and more benefits could be obtained by using a cavity with a higher loaded quality factor \cite{ahn2019high}.

\begin{figure}[tp]
	\centering
	\includegraphics[width=0.99\linewidth]{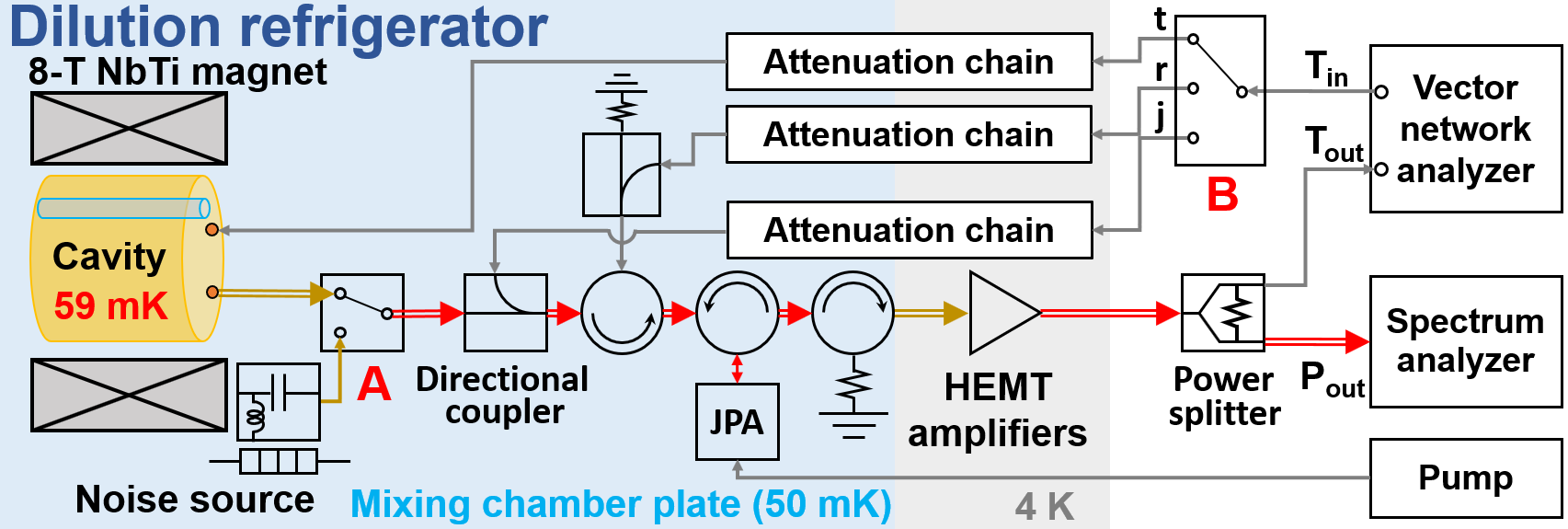}
	\caption{Schematic of the JPA-implemented CAPP-PACE detector. 
	At the 4K stage, two HEMT amplifiers with 1.5 K and 2.0 K of noise, respectively, are installed in series. A noise source capable of precise temperature control is connected to one side of switch~A, enabling a noise measurement of JPA and HEMT. Switch~B connects the network analyzer's input signal and the ports for measuring cavity transmission (t), cavity reflection (r), and JPA characteristics (j). 
	The arrows colored in yellow indicate NbTi superconducting line, while the rest of the lines were made of Cu or CuNi. 
	The arrows with double lines show the mainstream chain.}
\label{scheme}
\end{figure}

The CAPP-PACE consists of a 1.12-L split-style copper cavity with a sapphire tuning rod, an 8-T NbTi superconducting magnet, and a cryogenic RF receiver chain with a flux-driven JPA, a major upgrade from the previous experiment \cite{kwon2021first}. It is critical to design the RF receiver chain so that it has the lowest possible noise temperature, as the scan speed is proportional to the inverse-square of the system noise \cite{krauss1985calculations}. Figure~\ref{scheme} shows the new receiver chain with the JPA \cite{flux_jpa, JPA_squeezing, uchaikin2019development, capp_jpa, Nakamura_jpa} as a preamplifier and the associated electronics. The JPA was sealed with a two-layer superconducting magnetic shield and the shielded JPA was placed at the center of the mixing-chamber plate (MXC) of a dilution refrigerator, where the fringe magnetic field is the lowest ($\approx$ 200 G). The magnetic flux penetration was less than 0.5 magnetic flux quanta, and the JPA gain was maintained with a fluctuation of less than 0.1 dB during the experiment.

The system noise of the haloscope receiver chain with JPA can be expressed as \cite{capp_jpa, noise_figure, kwon2021first}.
\begin{align}
    \label{Tsys}
    T_\mathrm{sys} &=  T_\mathrm{phy} + T_\mathrm{add} \\
    T_\mathrm{add} &= (T_\mathrm{JPA,i} + T_\mathrm{JPA,e}) + \frac{T_\mathrm{HEMT}}{G_\mathrm{JPA}}
\end{align}
$T_\mathrm{phy}$ is the quantum and the thermal noise of the cavity, defined as $T_\mathrm{phy} = h\nu / k_B [1/(e^{h\nu/k_BT_\mathrm{cav}}-1) + 1/2]$, and $T_\mathrm{add}$ is the total added chain noise to $T_\mathrm{phy}$. $G_\mathrm{JPA}$ and $T_\mathrm{JPA}$ are the gain and the noise of the JPA, respectively. JPA noise $T_\mathrm{JPA}$ consists of an idler term ($T_\mathrm{JPA,i}$) and an extrinsic term ($T_\mathrm{JPA,e}$). $T_\mathrm{JPA,i}$ is the irreducible noise in the phase-insensitive operation and defines the quantum limit \cite{caves1982quantum}. It is theoretically expected to be identical to $T_\mathrm{phy}$ multiplied by the attenuation between the cavity antenna and the JPA ($\approx 1$ dB) \cite{noise_figure, dicke1946measurement, kwon2021first}, if the cavity and MXC are in thermal equilibrium. $T_\mathrm{JPA,e}$ is a collection of various noises that are generated by JPA operations. $T_\mathrm{HEMT}$ represents the noise generated in the post-JPA chain, where the HEMT noise is dominant.

In the cryogenic environment, the temperature of the MXC was controlled to be around 50 mK, achieving the stable operation of the JPA. The temperature of the cavity ($T_\mathrm{cav}$) was 59 mK under the 8-T magnetic field, producing about 75 mK of physical noise ($T_\mathrm{phy}$) in the frequency range near 2.3 GHz \cite{johnson1928thermal, nyquist1928thermal}. $T_\mathrm{HEMT}$ was measured to be 1.66 $\pm$ 0.06 K using the conventional Y-factor method \cite{engen1970new, yfactor, kwon2021first} with the inactivated JPA.

\begin{figure}[htp]
	\centering
	\includegraphics[width=0.98\linewidth]{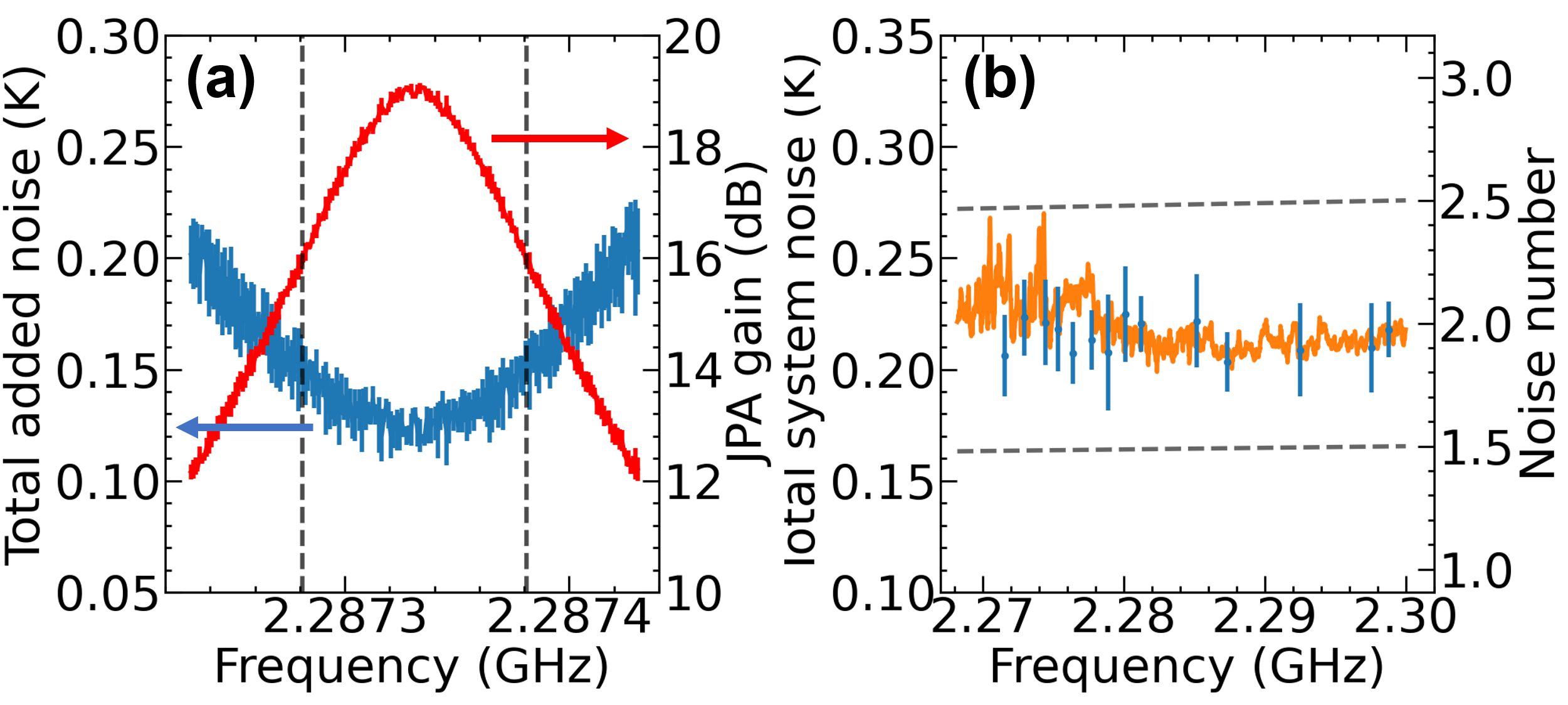}
	\caption{(a) Measurement of the receiver chain noise. The blue curve is the total added noise and the red curve is the JPA gain. The black dashed lines indicate the frequency window of the analysis (100~kHz). (b) Crosscheck between the noise tracked during Run~A (orange solid line) and the reference noise using the modified Y-factor method (blue dots). The tracked noise was obtained from the scan's power spectra. The black dashed line is a contour of the noise number, which is a frequency-dependent value.}
	\label{noiseTrack}
\end{figure}

To verify the JPA characteristics ($T_\mathrm{JPA}$ and $G_\mathrm{JPA}$), a modified Y-factor method was applied at three input noise power levels (60 mK, 120 mK, and 180 mK). The added noise and the system gain were determined via fitting to the JPA-specific equation, including the dependency of $T_\mathrm{JPA,i}$ on $T_\mathrm{phy}$ \cite{capp_jpa}. Figure~\ref{noiseTrack}(a) shows $T_\mathrm{add}$ and $G_\mathrm{JPA}$ of the receiver chain when the JPA resonant frequency was 2.2873 GHz and the JPA center gain was 19 dB. The estimated $T_\mathrm{add}$ at resonance was 129 $\pm$ 16 mK, and the system noise $T_\mathrm{sys}$ was correspondingly estimated to be 204 mK (see Table \ref{noiseContribution}). The extrinsic term of the JPA noise, $T_\mathrm{JPA,e}$, was kept small ($\approx 13$ mK), and the number of noise photons \cite{caves1982quantum} throughout the receiver chain was below 2 quanta. The JPA instantaneous bandwidth $\Delta f_\text{JPA}$ was $\sim$100 kHz, which was wide enough to cover the cavity bandwidth. The measurement was repeated at various JPA resonant frequencies, by tuning the frequency with the DC magnetic flux. While tuning, $\Delta f_\text{JPA} \geq 80$ kHz and $G_\text{JPA}\geq16$ dB were maintained by adjusting the pump tone. The noise measurement results in 2.27 -- 2.30 GHz range are plotted in Fig.~\ref{noiseTrack}(b)~(blue dots).

\begin{table}
\begin{ruledtabular}
\begin{tabular}{lccc}
\textrm{Noise contribution}&
\textrm{}&
\textrm{Measurement at 2.2873 GHz}\\
\colrule
$T_\mathrm{phy}$ & & 75 $\pm$ 3 mK ($0.69\times h\nu$)\\
$T_\mathrm{HEMT}/G_\mathrm{JPA}$ & & 21 $\pm$ 1 mK ($0.19\times h\nu$)\\
$T_\mathrm{JPA}$ ($=T_\mathrm{JPA,i}$+$T_\mathrm{JPA,e}$) & & 108 $\pm$ 15 mK ($0.99\times h\nu$)\\
\hline
$T_\mathrm{sys}$ (total) & & 204 $\pm$ 16 mK ($1.87\times h\nu$) \\
\end{tabular}
\end{ruledtabular}
\caption{Noise contributions of the RF chain characterized at 2.2873~GHz. $T_\mathrm{phy}$, $T_\mathrm{HEMT}$, and $T_\mathrm{add}$ were measured directly, while $T_\mathrm{JPA}$ was decoupled from $T_\mathrm{add}$ by calculation. The total system noise $T_\mathrm{sys}$ was 204 mK, which corresponds to 1.87 quanta of the noise number. The dominant part of the errors in $T_\mathrm{HEMT}$ and $T_\mathrm{JPA}$ was from the fittings in the Y-factor method.}
\label{noiseContribution}
\end{table}

Axion dark matter physics data were collected for the corresponding frequency range, with the parameters shown in Table \ref{runProps} utilizing the data acquisition process described in a previous work \cite{lee2017development, kwon2021first}. There were two experimental runs, Run~A and Run~B. From both runs a total of 4,940 power spectra were collected, which covered the 2.2704 -- 2.3000 GHz frequency range (corresponding to 9.39 -- 9.51 $\mu$eV in mass units). The span size of the spectra was set to 2~MHz to cover the entire cavity resonance region, and the resolution bandwidth was set to 100~Hz for the multi-bin search so that the axion signal shape ($\sim 1$ kHz) was properly resolved. The frequency step between the spectra in each run was 12~kHz, which was about 1/6 of the cavity frequency bandwidth. The cavity frequency tuning was accomplished by rotating a sapphire rod through a piezoelectric actuator, followed by the tuning of the JPA's resonant frequency. Since this process changed the internal state of the JPA, the power level of the spectrum and the gain of the system were examined to track the system noise. The change in the system noise from the reference frequency (2.3 GHz) were compared, where the noise at the reference frequency was measured accurately using the modified Y-factor method. The average tracked system noise of Run~A was found to be 220~mK ($2.00 \times h\nu$), which was consistent with the noise values from the Y-factor measurement [see Fig.~\ref{noiseTrack}(b)]. The average system noise in Run~B increased to 260~mK ($2.35\times h\nu$), mainly due to the poor thermal contact between the cavity and the MXC plate.

\begin{table}
\begin{ruledtabular}
\begin{tabular}{lcc}
\textrm{Experimental runs}&
\textrm{Run~A}&
\textrm{Run~B}\\
\colrule
Period (2020) \footnote{The time for maintenance is included.} & Apr 01 -- May 11  & Sep 05 -- Sep 18 \\
Frequency range (GHz) & 2.2704 -- 2.3000 & 2.2705 -- 2.2997 \\
Axion mass $m_a$ ($\mu$eV) & 9.3906 -- 9.5130 &9.3910 -- 9.5118\\
Frequency bin width & \multicolumn{2}{c}{100 Hz}\\
Single tuning step & \multicolumn{2}{c}{12 kHz}\\
Sweep time $\tau_0$ \footnote{Pure sweep time without dead time} & \multicolumn{2}{c}{60 sec}\\
Number of spectra \footnote{Number of spectra per step} & 7 & 5 \\
Sweep time per step $\tau$ \footnote{$\tau =$ Number of spectra per step $\times \tau_0$} & 7 min & 5 min\\
Number of steps & 2,485 & 2,455 \\
Quality factor $Q_0$ \footnote{$Q_0$ degradation comes from the oxidization of the cavity.} & 90,000 & 86,000 \\
Magnetic field & \multicolumn{2}{c}{7.2 T} \\
Cavity volume &  \multicolumn{2}{c}{1.12 L} \\
Geometrical factor  & \multicolumn{2}{c}{0.45} \\
MXC temperature & \multicolumn{2}{c}{50 mK} \\
Cavity temperature & 59 mK & 79 mK \\
System noise $T_\mathrm{sys}$ \footnote{Noise values are averaged over total frequency range.} & 220 mK & 260 mK \\
\end{tabular}
\end{ruledtabular}
\caption{Major parameters of two experimental runs.}
\label{runProps}
\end{table}

The analysis procedure was essentially the same as in the previous experiments \cite{ADMX_PRL_1998, brubaker2017first, lee2020axion, *jeong2020search, kwon2021first}. The analysis began with the removal of the baseline in the power spectrum, which contained the effect of the frequency-dependent gain \cite{capp_jpa} and impedance mismatch of the receiver chain \cite{PRD_ADMX2001}. The shape of the power spectrum baseline was estimated using the Savitzky-Golay (SG) filter \cite{sgfilter}, and extracting it from the original spectrum restored the normal distribution of the noise profile. Conducting an optimization of the filter parameters \cite{brubaker2017haystac} with a Monte-Carlo simulation, the signal reduction due to the baseline removal was minimized to around 20\%, when the polynomial order $d$ was 5 and the window length $2W+1$ was 391. Next, a process called ``combination"~\cite{lee2020axion, *jeong2020search} completed the construction of a grand spectrum over the entire scan range by convolving the noise-normalized axion line shape and the baseline-removed data \cite{turin1960introduction}. The axion velocity dispersion with respect to the galactic center was set to 270~km/s, and 230~km/s was used for the orbital speed of the Sun around the galaxy \cite{turner1990periodic}.

The final step in the analysis was to define and examine axion candidates using a threshold-based criterion, where a threshold for axion candidates was defined using a signal-to-noise ratio (SNR) and a confidence level. Conventionally, once a target sensitivity level $g_{a\gamma\gamma}$ and a target SNR (usually 5) are set, the total scan time $\Delta t_\mathrm{total}$ can be determined based on the experimental parameters \cite{lee2020axion, brubaker2017first}. To reduce $\Delta t_\mathrm{total}$, we developed a two-stage scanning method which splits the conventional scan into two scans, and inserts an intermediate threshold between scans. The initial scan first sweeps the entire frequency range with the preset $g_{a\gamma\gamma}$, and the intermediate SNR less than 5. A second scan is then performed over those surviving frequency bins. The second scan data are collected with enough integration time to make the final SNR become 5, when merged with the initial scan. From the combined data, axion candidates are determined using the final threshold of SNR 5. In this way, unlike the conventional scan, not all frequency bins are investigated until the SNR reaches 5, which reduces the total scan time without changing the final confidence level.

\begin{figure}[t]
	\centering
	\includegraphics[width=0.98\linewidth]{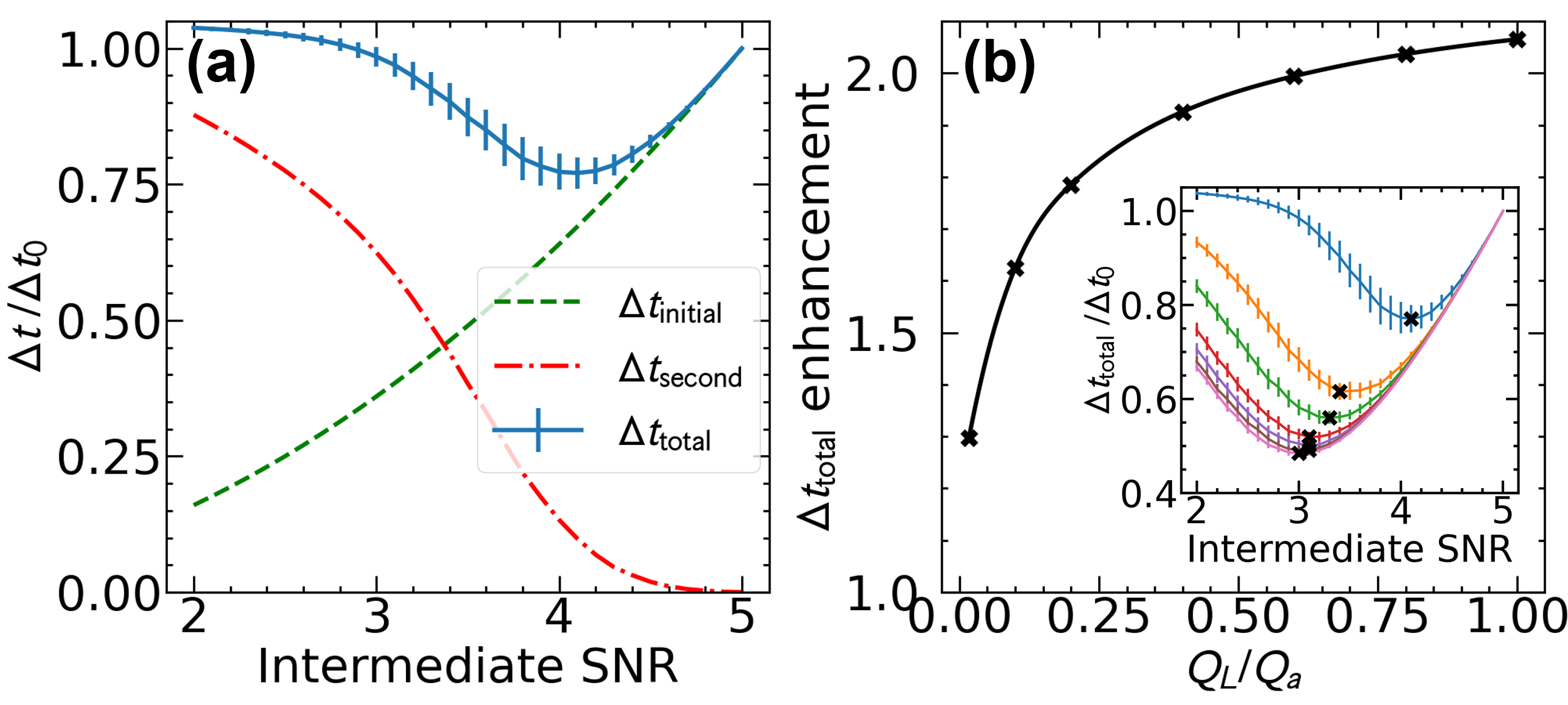}
	\caption{
	(a) Simulation of the total scan time $\Delta t_\mathrm{total}$ as a function of the intermediate SNR, when applying a two-stage scanning method with a cavity that has a loaded quality factor of 30,000. The time unit is normalized to the scan time of the conventional scan, $\Delta t_0$, and $\Delta t_\mathrm{total}$ is given as a sum of $\Delta t_\mathrm{initial}$ and $\Delta t_\mathrm{second}$. The error bars indicate statistical errors by the stochastic behavior of the second scan. (b) Enhancement in the scan speed with respect to the loaded quality factor of the cavity, $Q_L$, which is normalized to the quality factor of the axion, $Q_a$. The inset shows the simulation results using the same method as in (a), but with various $Q_L$ values. The markers in the inset indicate the optimized points that correspond to the markers in the main plot.}
	\label{snr_study}
\end{figure}

\begin{figure*}[htp]
	\centering
	\includegraphics[width=0.99\linewidth]{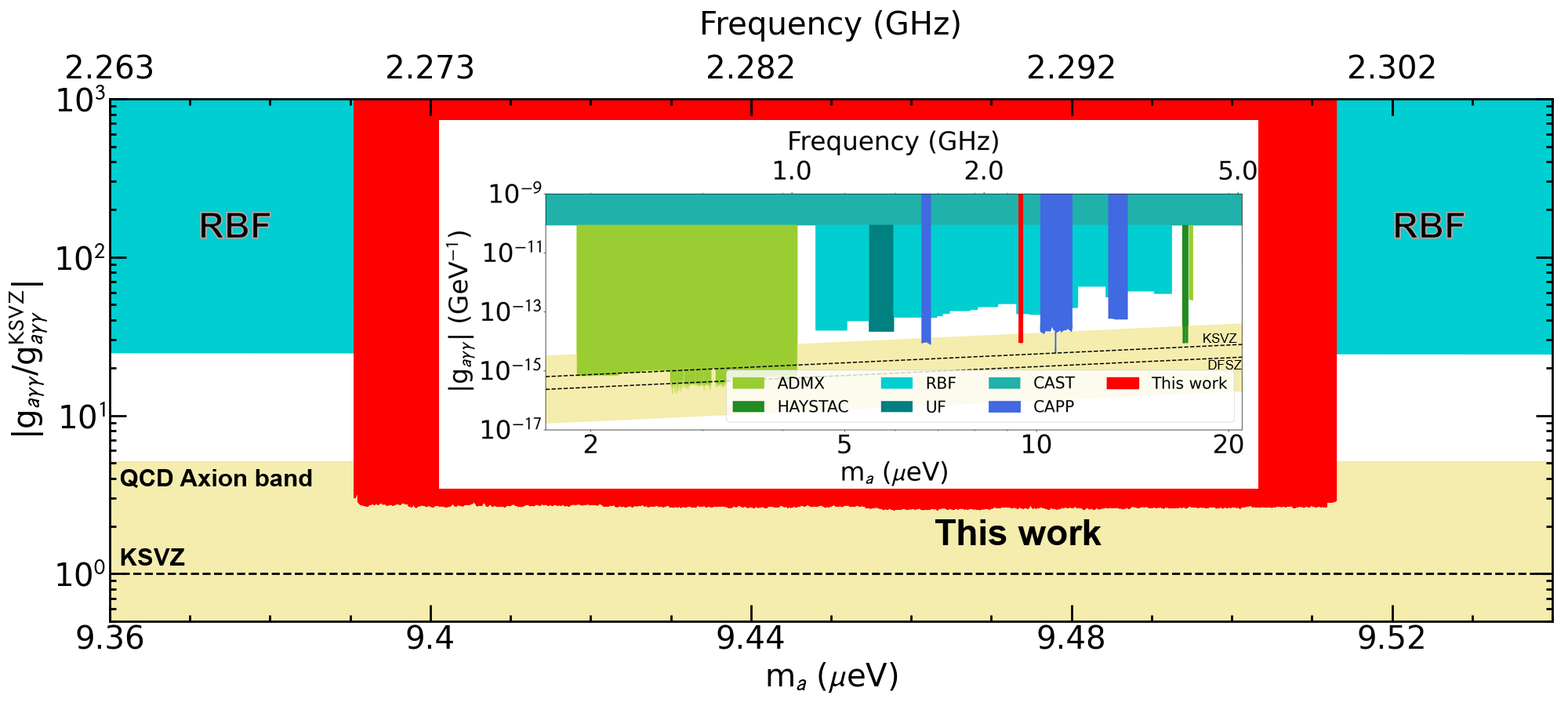}
	\caption{Exclusion limit of this work at 90\% confidence (red area). The inset shows this work along with other axion search results in the extended axion mass range \cite{RBF1, RBF2, UF, ADMX_PRL_1998, Asztalos_2002, PRD_ADMX2004, PRL_ADMX2010, PRL_ADMX2018, PRL_ADMX2018_2, PRL_ADMX2020, PRL_ADMX2021, PRD_cast_2015, cast_nature, zhong2018results, haystac_nature2021}. The results from CAPP except this work are colored blue \cite{lee2020axion, *jeong2020search, kwon2021first}. The yellow area shows the model band of QCD axion \cite{peccei2008strong, cheng1995axion}.}
	\label{exclusion}
\end{figure*}

To determine the optimal value of the intermediate SNR and minimize the total scan time, a Monte-Carlo simulation was conducted with various intermediate SNR values. Figure~\ref{snr_study}(a) shows how much scan time could be saved by having an intermediate SNR when $g_{a\gamma\gamma}$ was fixed. The final confidence level was controlled to 90\%, while the confidence level at each scan was set slightly higher than 90\%. Meanwhile, the efficiency of the second scan increased as the loaded quality factor of the cavity, $Q_L$, approached the axion quality factor $Q_a$, since the cavity with high $Q_L$ has a bandwidth that can concentrate more on the surviving bins in the second scan. The enhancement of $\Delta t_\mathrm{total}$ with increasing $Q_L$ is presented in  Fig.~\ref{snr_study}(b). For the cavity used in this experiment [blue in Fig.~\ref{snr_study}(a) and (b)], $\Delta t_\mathrm{total}$ was found to be at minimum when the intermediate SNR was around 4.1, and the enhancement in scan speed was as much as $\sim$30\%. If the superconducting cavity \cite{ahn2019high} is utilized so that $Q_L$ matches $Q_a$, the scan speed could be boosted up to 105\%.

In the analysis of this experiment, the combined data of Run~A and Run~B formed the initial scan, and the optimal intermediate SNR for this experiment was found to be 4.2. There were a total of 1,168 surviving frequency bins from the initial scan, which is close to the analytically expected number. When the data from the initial and the second scans were combined according to the two-stage scanning method, 46 axion candidates were found from the surviving bins. After a re-scan with sufficiently high statistics, the candidates were completely excluded with the 90\% confidence level. The actual improvement in scan speed by applying the two-stage scanning method was found out to be around 26\%, similar to the simulation result. Assuming that the axion portion of dark matter for our local galactic halo is 100\%; i.e., $\rho_a = 0.45 \ \mathrm{GeV/cm^3}$, we obtained a coupling sensitivity of 2.7-fold above the Kim-Shifman-Vainshtein-Zakharov (KSVZ) level (see Fig.~\ref{exclusion}).

In conclusion, we report the first result of an axion haloscope search using a quantum-noise-limited JPA at CAPP. By implementing the JPA in the cryo-RF receiver chain, it was possible to reduce the total system noise down to approximately 200~mK in the 2.27 -- 2.30~GHz frequency region, which is less than twice the standard quantum limit. We were able to achieve the lowest system noise recorded among published axion cavity haloscope searches under phase-insensitive operations, in terms of both the absolute temperature and the noise photon number. Thanks to the new receiver chain, the scan speed was improved by about a factor of 50 compared to the case with an inactivated JPA. In addition, a new scanning method was developed, which boosted the scan speed by $\sim$26\%. It is expected that this can be enhanced even more when a superconducting cavity is implemented in future work. As a result, we were able to exclude the
$9.39 < m_a < 9.51$ $\mu$eV axion mass range with an order higher sensitivity than the existing limit,
where the parameters are in the plausible model for the QCD axion \cite{peccei2008strong, cheng1995axion}.

\begin{acknowledgments}
\indent This work is supported in part by the Institute for Basic Science (IBS-R017-D1-2022-a00) and JST ERATO (Grant No.~JPMJER1601). Arjan Ferdinand van Loo was supported by a JSPS postdoctoral fellowship.
\end{acknowledgments}



\bibliography{reference}

\end{document}